\documentclass{article}
\usepackage{amsmath,amssymb,amsthm,amscd,latexsym,epsfig,subfig}
\usepackage[text={6.5in,9.in},centering]{geometry}

\newcommand{\be}{\begin{equation}}
\newcommand{\ee}{\end{equation}}

\newenvironment{pf}{\textbf{Proof:}\quad}
{\nopagebreak{\begin{flushright}{$\blacksquare$}\end{flushright}}}

\renewcommand{\O}{\mathcal{O}}

\newtheorem{thm}{Theorem}

\newtheorem{defn}{Definition}
\newtheorem{remark}{Remark}

\numberwithin{equation}{section}
\numberwithin{lem}{section}

\newcommand{\e}{\epsilon}

\newcommand{\bu}{\bar{u}}

\begin{document}
\title{Anomalous spreading in a system of coupled Fisher-KPP equations}

\author{Matt Holzer\footnote{mholzer@gmu.edu, 612-625-1119}\footnote{current address: Department of Mathematics, George Mason University, Fairfax, VA 22030} \\
University of Minnesota \\
School of Mathematics \\
127 Vincent Hall, 206 Church St SE \\
Minneapolis, MN 55455, USA}

\date{ \today}

\maketitle
\begin{abstract}
In this article, we report on the curious phenomena of anomalous spreading in a system of coupled Fisher-KPP equations.  When a single parameter is set to zero, the system consists of two uncoupled Fisher-KPP equations which give rise to traveling fronts propagating with the unique, minimal KPP speed.  When the coupling parameter is nonzero various behaviors can be observed.  Anomalous spreading occurs when one component of the system spreads at a speed significantly faster in the coupled system than it does in isolation, while the speed of the second component remains unchanged.  We study these anomalous spreading speeds and show that they arise due to poles of the pointwise Green's function corresponding to the linearization about the unstable homogeneous state.  These poles lead to anomalous spreading in the linearized system and come in two varieties -- one that persists and leads to anomalous spreading for the nonlinear system and one that does not.  We describe mechanisms leading to these two behaviors and prove that one class of poles are irrelevant as far as nonlinear wavespeed selection is concerned.  Finally, we show that the same mechanism can give rise to anomalous spreading even when the slower component does not spread.

\end{abstract}



\section{Introduction}
In this article, we study anomalous spreading speeds in a system of coupled Fisher-KPP equations,
\begin{eqnarray}
u_t &=& du_{xx}+\alpha\left(u-u^2\right)+\beta v (1-u) \nonumber \\
v_t &=& v_{xx}+\left(v-v^2\right) \label{eq:main}.
\end{eqnarray}
We are interested in the following problem.  Consider a positive, compactly supported perturbation of the unstable homogeneous state $(u,v)=(0,0)$.  What is the asymptotic speed of propagation associated to the $u$ component?  

\begin{figure}[ht]
\centering
   \includegraphics[width=0.4\textwidth]{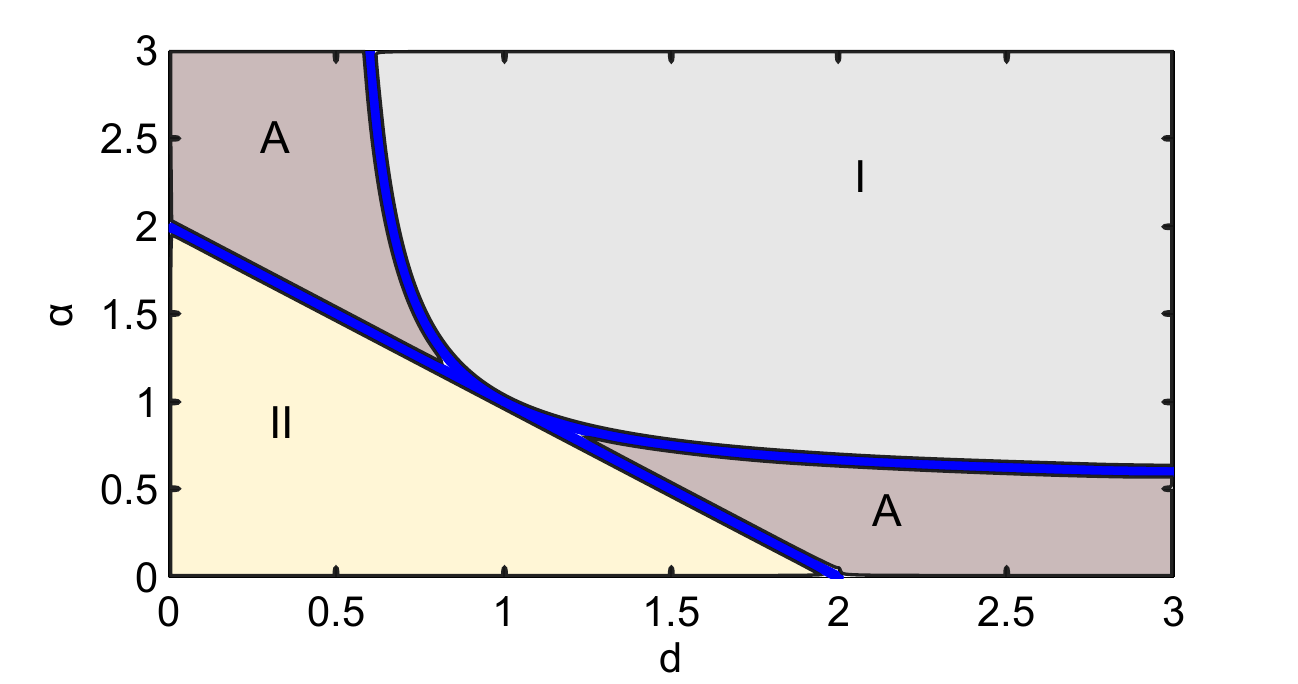}
   \includegraphics[width=0.4\textwidth]{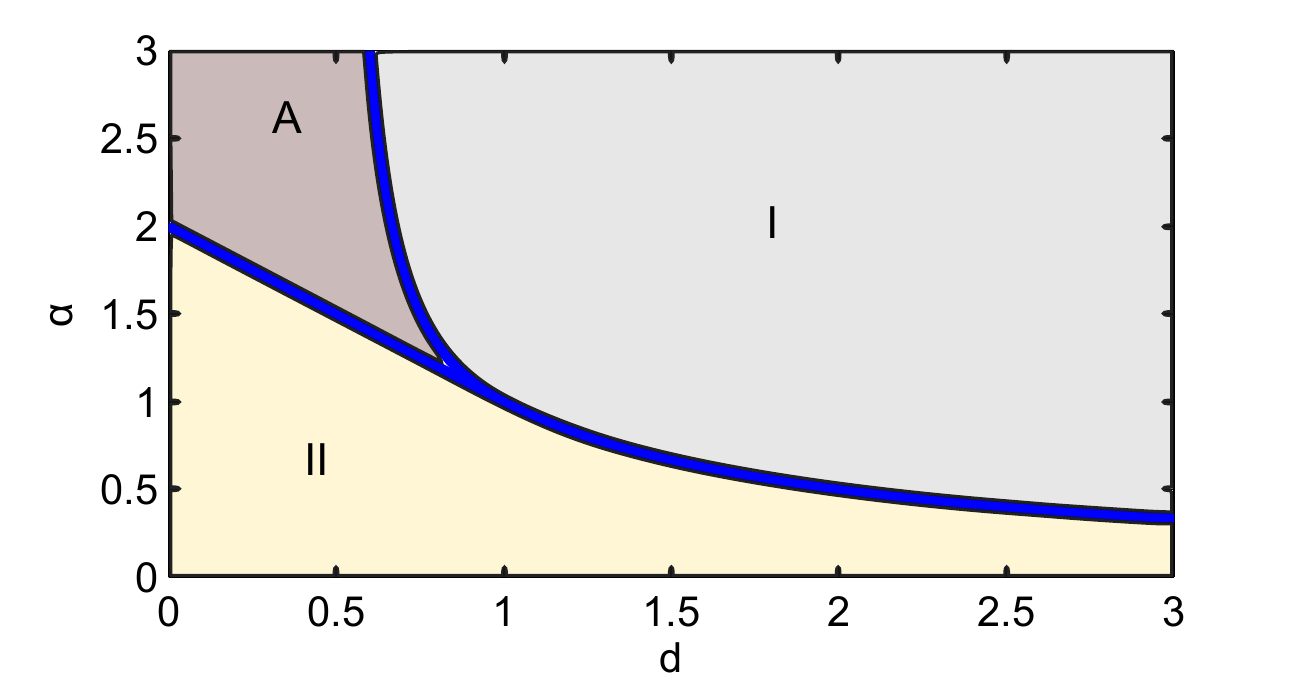}  
\caption{The selected spreading speed for the linear (left) and nonlinear (right) systems.  The linear speeds are calculated explicitly while the nonlinear speeds are measured from numerical simulations.  Here $I$ denotes a region where the selected speed of the $u$ component is the pulled speed $2\sqrt{d\alpha}$, whereas in region $II$ the selected speed is $2$.  The region $A$ is where anomalous speeds are observed.  Note the difference between the linear and nonlinear cases - the anomalous speeds arising due to fast diffusion of the $u$ component ($d>1$) do not persist in the nonlinear regime.}
\label{fig:parameterspace}
\end{figure}

Consider first the case when $\beta=0$ or $v$ is identically zero.  Then the equation governing the dynamics of the $u$ component is the scalar Fisher-KPP equation,
\[ u_t=du_{xx}+\alpha(u-u^2).\]
This equation has been studied in great detail by a number of authors, see \cite{fisher37,kolmogorov37,aronson78,bramson83} among others.  It is well known that compactly supported, positive perturbations of the zero state evolve into a pair of counter propagating fronts.  These fronts travel with asymptotic speed $2\sqrt{d\alpha}$.  At the same time, observe that the $v$ component decouples and its evolution is also given by a Fisher-KPP equation with a selected spreading speed of two.  For $\beta>0$ the evolution of the $u$ component depends on that of the $v$ component.  One might conjecture that the selected speed of the $u$ component for the full system will be either $2$ or $2\sqrt{d\alpha}$, whichever is larger.  This turns out to not always be the case, and there are large swaths of the $(d,\alpha)$ parameter plane for which speeds of propagation are observed that exceed both $2$ and $2\sqrt{d\alpha}$.  See the right panel of Figure~\ref{fig:parameterspace} for an illustration of those parameter values leading to faster speeds of propagation in the $u$ component.  This phenomena was first observed in \cite{weinberger07} and given the label of anomalous spreading.  In this article, we give a complete description of the anomalous spreading that arises in system (\ref{eq:main}) and discuss the mechanisms leading to these faster than expected speeds of propagation.  

A natural starting point for the analysis is the linearization of (\ref{eq:main}) about the homogeneous unstable zero state.  The fastest speed of propagation associated to the linearization is called the linear spreading speed and is of interest for many reasons.  Practically speaking, the linear spreading speed can be inferred from the singularities of the pointwise Green's function.  We delay a precise description of this process until section~\ref{sec:linear} but we remark that this was first observed within the plasma physics community; see \cite{briggs,bers84} for descriptions of the original work aimed at differentiating between absolute and convective instabilities.  

From a physical standpoint, one is generally interested not in the linear spreading speed but in the nonlinear spreading speed.  One way in which the linear spreading speed is significant is as a predictor for the spreading speed of the nonlinear system.  This prediction is often reliable, and fronts propagating with the linear spreading speed are referred to as pulled fronts according to the commonly used terminology reviewed for example in \cite{vansaarloos03}.  Examples of pulled fronts abound and we refer to \cite{vansaarloos03} for a large number of examples.  Of course, the nonlinear spreading speed often differs from the linear spreading speed.  The canonical example here is Nagumo's equation, see \cite{hadeler75}, where the selected nonlinear spreading speed exceeds the linear spreading speed.  These nonlinearly determined fronts are referred to as pushed fronts.  The study of wavespeed selection and invasion fronts represents a large and interesting area of research, see \cite{xin,vansaarloos03} for review articles.  In the context of cooperative reaction-diffusion equations we mention \cite{lewis02,li05} where a large class of cooperative systems were shown to be linearly determinate.  Other studies of wavespeed selection in systems of reaction-diffusion equations include studies related to autocatalytic chemical reactions \cite{billingham91,focant98},  combustion \cite{berestycki85},  competitive population models \cite{hosono03}, $\lambda-\omega$ type equations \cite{sherratt09}, phase field equations \cite{goh11}, and staged invasion processes \cite{holzerAF} among many others.

Another natural question is whether the linear spreading speed places a lower bound on the speed of propagation for the nonlinear system.  Wavespeed selection for general systems of equations is usually described in terms of marginal stability of the invasion front, \cite{dee83,vansaarloos88}.  The selected front should be marginally stable against compactly supported perturbations, which is to say that perturbations of the traveling front should neither grow nor decay when viewed pointwise in a traveling frame moving with the speed of that front.  Since the linear spreading speed describes the spreading speed of perturbations of the unstable homogeneous state, one might imagine that a perturbation placed sufficiently far ahead of the front interface should spread with the linear spreading speed.  This seems to suggest that the linear spreading speed should place a lower bound on the speed of propagation for the nonlinear system.  This is the case for scalar problems, but for systems this bound no longer holds.  This fact was recently shown in an example of a Lotka-Volterra competition model, \cite{holzerLV},
\begin{eqnarray}\label{eq:lotka}
u_t&=& \e^2 u_{xx}+(1-u-a_1v)u \nonumber \\
v_t&=& v_{xx}+r(1-a_2u-v)v.
\end{eqnarray}
Here $0<\e\ll 1$, $r>0$ and $a_1<1<a_2$.  The linear spreading speed in this example is anomalous and $\O(1)$ while the nonlinear spreading speed is $\O(\e)$, an order of magnitude slower.  This should be contrasted with the dynamics of system (\ref{eq:main}) as well as the example provided in \cite{weinberger07}.  In both those cases, anomalous spreading speeds are observed in both the linear and nonlinear systems.  Therefore, one aim of this article is to determine how anomalous linear speeds persist to the nonlinear regime in some cases but not in others.  

Mathematically, we will show that these anomalous linear spreading speeds arise due to poles of the pointwise Green's function.  This is a crucial difference between the scalar and non-scalar cases.  Systems of reaction-diffusion equations can support poles of the pointwise Green's function whereas these singularities in the scalar case are always accompanied by a loss of analyticity.  Poles of the pointwise Green's function are important because they represent the linear spreading of one component that is induced by the coupling to a component spreading with a different speed.  Based upon the results depicted in Figure~\ref{fig:parameterspace} as well as the examples in \cite{weinberger07,holzerLV} it seems that these poles come in two varieties: one that induces faster spreading speeds in the nonlinear system and one that does not.  In section~\ref{sec:nonlinear}, we will come to this question and describe two different mechanisms leading to anomalous linear spreading speeds.  One such mechanism persists in the nonlinear system, while the second type does not.  We remark that for poles of the pointwise Green's function to exist, the linearization must possess a skew-product structure where some of the components decouple from the others.  Problems for which this skew-product structure exist arise in many areas of pattern formation, ecology and chemistry among others, see \cite{bell09,iida11,goh11,kovacs10,mccalla12} for recent examples.


That anomalous spreading exists despite the fact that the $v$ component is converging pointwise to zero in a frame moving with the anomalous spreading speed suggests that this spreading phenomena should not be dependent on the instability of the zero state with respect to perturbations of the $v$ component.  In fact, this is the case.  For example, suppose the $v$ component in (\ref{eq:main}) was altered so that the following system was under consideration,
\begin{eqnarray*}
u_t &=& du_{xx}+\alpha\left(u-u^2\right)+\beta v (1-u) \nonumber \\
v_t &=& v_{xx}- \gamma v 
\end{eqnarray*}
For the $v$ component, the zero state is stable and any perturbation of that state will relax to zero.  However, in the process of this relaxation $v(t,x)$ is non-zero and has certain decay rates as $x\to\infty$.  In an analogous manner to the anomalous spreading for (\ref{eq:main}), the effect of this spatial decay can give rise to faster speeds of propagation in the nonlinear system.  We give an explicit example of this in section~\ref{sec:stablebutfast}.  An important conclusion to be drawn here is that one must use caution when reducing a system of reaction-diffusion equations by setting stable or slowly propagating components to some constant value.  The reduced system have have very different dynamics than the original.

The paper is outlined as follows.  In section~\ref{sec:linear} we compute the linear spreading speed using the pointwise Green's function.  In section~\ref{sec:nonlinear}, we discuss under what circumstances anomalous spreading in the linearized equation leads to anomalous nonlinear spreading.  In section~\ref{sec:stablebutfast}, we give an example that shows that this phenomena arises even when the component that induces the anomalous spreading is pointwise stable.  In section~\ref{sec:numerics}, we contrast numerically observed spreading speeds with the predictions based upon the linearized system.  We conclude in section~\ref{sec:discussion} with a short discussion.

\section{Spreading speeds in the linearized system}~\label{sec:linear}
In this section, we will compute the linear spreading speed associated to the $u$ component in (\ref{eq:main}).  That is, we will linearize (\ref{eq:main}) about the unstable zero state and compute the asymptotic speed of propagation associated to compactly supported perturbations of the unstable state.  This linear speed will give a prediction for the observed spreading speeds in the nonlinear system.  We will return to the question of the reliability of this prediction in section~\ref{sec:nonlinear} and concern ourselves with only the linear system in this section.

The linear spreading speed can be thought of intuitively as the infimum of all speeds for which a moving observer will outrun the instability.   Put another way, the linear spreading speed is the largest speed for which a transition from absolute to convective instability is observed.  This transition can be naturally understood in terms of singularities of the pointwise Green's function.  After linearizing system (\ref{eq:main}) about the unstable zero state, we will compute these singularities explicitly and calculate the linear spreading speed.  
\paragraph{Preliminaries - linear spreading speeds, the pointwise Green's function and pinched double roots of the dispersion relation}

To begin, we linearize (\ref{eq:main}) about the unstable zero state and consider its evolution in a moving coordinate frame.  Let $\xi=x-st$ for some $s>0$.  Then the linearization in this coordinate frame is
\begin{eqnarray}
u_t &=& du_{\xi\xi}+su_\xi+\alpha u+\beta v \nonumber \\
v_t &=& v_{\xi\xi}+sv_\xi+v \label{eq:mainlinearized}.
\end{eqnarray}
Since the zero state is unstable, compactly supported perturbations of this state will grow in norm (say the $L^2(\mathbb{R})$ norm) and spread spatially.  To capture the speed of this spreading, we are interested in the pointwise growth rate of the solution as a function of the wavespeed $s$.  In particular, we seek the value of $s$ for which a transition from pointwise decay to pointwise growth is observed.  This transition is understood as a transition from absolute to convective instability.  To recall, an absolute instability is one wherein the value of the solution grows exponentially for each fixed value of $\xi$.  When the instability is convective, the solution again grows in the $L^2$ norm but is transported away from its original location and pointwise decay to zero is eventually achieved despite exponential growth in norm.  The question of how to mathematically differentiate between convective and absolute instabilities has a long history in the literature, see \cite{briggs,bers84,huerre90,brevdo96,sandstede00,vansaarloos03} for the original work within the plasma physics community and subsequent extensions.  The key observation is that absolute and convective instabilities can be distinguished according to whether the pointwise Green's function, $G_\lambda(\xi-y)$, has singularities in the right half plane.  

We review some basic information about the pointwise Green's function $G_\lambda(\xi-y)$ pertinant for the study (\ref{eq:mainlinearized}).  We refer the reader to \cite{zumbrun98} for a more detailed treatment in the context of viscous shock waves.  The pointwise Green's function is computed as follows.  After a Laplace transform in time, the linear system (\ref{eq:mainlinearized}) reduces to a system of second order ordinary differential equations with inhomogeneous right hand side.  These equations depend on $\lambda$ in an analytic fashion and for each $\lambda$ the corresponding Green's function for this system of equations is denoted $G_\lambda(\xi-y)$ and satisfies
\be \left(\begin{array}{cc} \mathcal{L}_{u}-\lambda & \beta \\ 0 & \mathcal{L}_{v}-\lambda \end{array}\right)\left(\begin{array}{cc} G^{11}_\lambda & G^{12}_\lambda \\ G^{21}_\lambda & G^{22}_\lambda \end{array}\right)=\left(\begin{array}{cc} \delta(y) & 0 \\ 0 & \delta(y) \end{array}\right). \label{eq:Glambda} \ee
Let $\mathcal{L}-\lambda$ denote the matrix operator on the left-hand side of (\ref{eq:Glambda}), posed on $L^2(\mathbb{R})\times L^2(\mathbb{R})$.  The pointwise Green's function is the kernel of the resolvent operator via the identity
\[ [(\mathcal{L}-\lambda)^{-1}f](x)=\int_\mathbb{R} G_\lambda(x-y)f(y)dy.\]
To the right of the essential spectrum of $\mathcal{L}-\lambda$ the resolvent operator is analytic and the pointwise Green's function inherits this analyticity.  The solution of the original initial value problem can then be determined via inverse Laplace transform,
\[ \left(\begin{array}{c} u(t,x) \\v(t,x) \end{array}\right)=\frac{-1}{2\pi i}\int_\Gamma e^{\lambda t}\int_\mathbb{R} G_\lambda(x-y)\left(\begin{array}{c} u(0,y) \\v(0,y) \end{array}\right)dyd\lambda,\]
where $\Gamma$ is the Laplace inversion contour and must lie to the right of any singularities of $G_\lambda$.  This inversion contour may be deformed so long as $G_\lambda$ does not have a singularity.  In this way, bounds can be placed upon the solution of the initial value problem and pointwise decay is achieved if $G_\lambda$ can be analytically continued into the left half plane.  Of course, analyticity properties of $G_\lambda(\xi-y)$ depend on the frame of reference, or on the value of the wavespeed $s$.  For $s=0$, $G_\lambda(\xi-y)$ has a singularity in the right half plane located at the extremal point of the essential spectrum in the $\lambda$ plane.  As $s$ is increased the singularities of $G_\lambda$ move until eventually some value $s_{lin}$ is reached so that all singularities of $G_\lambda(\xi-y)$ remain in the stable half plane for all $s>s_{lin}$.  This value of the wavespeed at which the singularity crosses the imaginary axis is the linear spreading speed.  We have the following definition. 
\begin{defn}
The linear spreading speed is 
\[ s_{lin}=\sup_{s>0}\left\{ G_\lambda(\xi-y) \ \ \text{has a singularity for some} \ \lambda \ \text{with} \ \mathrm{Re}(\lambda)>0\right\}. \] 
\end{defn}

In our example, singularities of the pointwise Green's function are related to the eigenvalues of the operators $\mathcal{L}_{u,v}-\lambda$.  Denote these eigenvalues $\nu_u^\pm(\lambda,s)$ and $\nu_v^\pm(\lambda,s)$ where we will henceforth suppress the dependence on $\lambda$ and $s$.  These eigenvalues are roots of the characteristic polynomials, $d_u(\nu,\lambda)=d\nu^2+s\nu+\alpha-\lambda$ and $d_v(\nu,\lambda)=\nu^2+s\nu+1-\lambda$.  The product of these polynomials is the dispersion relation, 
\[ d(\lambda,\nu)=d_u(\nu,\lambda)d_v(\nu,\lambda)=(d\nu^2+s\nu+\alpha-\lambda)(\nu^2+s\nu+1-\lambda), \]
roots of which relate spatial modes $e^{\nu x}$ to their temporal growth rate $e^{\lambda t}$.  For $\lambda$ lying to the right of $\sigma_{ess}(\mathcal{L})$, the eigenvalues $\nu^+_{u,v}$ lie to the right of the imaginary axis while the stable eigenvalues $\nu_{u,v}^-$ lie to the left of the imaginary axis.  Analytic continuation of $G_\lambda$ depends on the ability to analytically continue the stable and unstable subspaces associated to these eigenvalues into the essential spectrum.  Potential issues arise at values of the spectral parameter $\lambda$ for which an eigenvalue that originated on one side of the imaginary axis aligns with an eigenvalue that originated on the opposite side of the imaginary axis, indicating a non-trivial intersection of the analytical continuation of the stable and unstable subspaces.  The points at which these two roots coincide are referred to as {\em pinched double roots} of the dispersion relation. That is, $d(\lambda,\nu)$ has a pinched double root at $(\lambda^*,\nu^*)$ if 
\[ d(\lambda^*,\nu^*)=0, \quad \partial_\nu d(\lambda^*,\nu^*)=0 ,\quad Re\ \nu^\pm(\lambda)\to \pm \infty \ \text{as} \ \mathrm{Re}(\lambda)\to\infty.\]
We remark that while pinched double roots may prevent analytic continuation of $G_\lambda$, the existence of a pinched double root is not a sufficient condition for the existence of a singularity of $G_\lambda$.  This is because the pointwise Green's function has an analytic continuation if the stable and unstable subspaces can be continued analytically and pinched double roots do not imply that these subspaces are not analytic.  This is easy to observe in our example.  If $\beta=0$, the equations are decoupled and it is clear that a value of $\lambda$ for which $\nu_u^\pm=\nu_v^\mp$ should not lead to a singularity of $G_\lambda$.  Nonetheless, for $\beta\neq 0$ these pinched double roots do give rise to singularities of the pointwise Green's function.  We will now explicitly compute the pointwise Green's function and show that, in some instances, these pinched double roots induce poles of the pointwise Green's function and anomalous spreading speeds.

\paragraph{Spreading speeds from the dispersion relation}

Due to the skew-product nature of the linearization in (\ref{eq:mainlinearized}), we can compute the roots of the dispersion relation explicitly.  We find
\begin{eqnarray}
\nu_u^\pm(\lambda,s)&=& \frac{-s}{2d}\pm\frac{1}{2d}\sqrt{s^2-4d\alpha+4d\lambda} \nonumber \\
\nu_v^\pm(\lambda,s)&=& \frac{-s}{2}\pm\frac{1}{2}\sqrt{s^2-4+4\lambda}.\label{eq:nus}
\end{eqnarray}
Pinched double roots of the full dispersion relation occur whenever $\lambda$, $\nu$ and $s$ are such that  $\nu_{u,v}^+=\nu_{u,v}^-$.  Each pinched double root induces a spreading speed.  This spreading speed can be found by selecting $s$ in such a way that the value of $\lambda$ at the double root satisfies $Re(\lambda)=0$.  We set $\lambda=0$ and solve for the values of $s$ that give rise to pinched double roots.  The spreading speeds associated to the $u$ and $v$ components in isolation are easily found by setting the terms inside the roots in (\ref{eq:nus}) to zero.  A third spreading speed, the anomalous spreading speed is found by solving for values of $s$ for which $\nu_u^\pm=\nu_v^\mp$ with $\lambda$ again set equal to zero.  Doing this, we find the following three speeds
\begin{eqnarray*}
s_{u,lin}&:=&2\sqrt{d\alpha} \\
s_{v,lin}&:=&2 \\
s_{anom}&:=& \sqrt{\frac{\alpha-1}{1-d}}+\sqrt{\frac{1-d}{\alpha-1}}.
\end{eqnarray*}
Depending on what parameter values, $(d,\alpha)$, are under consideration the selected spreading speed of the $u$ component is one of these three speeds.  We have the following result.
\begin{thm}~\label{thm:linearss} The linear spreading speed for system (\ref{eq:mainlinearized}) is 
\be
s_{lin}=\left\{ \begin{array}{c}
 2 \ \  \text{for} \ \ \alpha<2-d  \\
2\sqrt{d\alpha} \ \ \text{for}\ \alpha>\frac{d}{2d-1}  \ \ \text{and} \ \ (d>\frac{1}{2})  \\
\sqrt{\frac{\alpha-1}{1-d}}+\sqrt{\frac{1-d}{\alpha-1}} \ \text{otherwise} \ 
\end{array}\right. \ee
\end{thm}
These regions are plotted in the left panel of Figure~\ref{fig:parameterspace}.  This result is established by verifying that the pinched double root calculation that leads to the derivation of these speeds actually induces a singularity of the pointwise Green's function.

\paragraph{Computing the pointwise Green's function}

We will now compute $G_\lambda$.  As is expected, this function will have singularities at the values of $\lambda$ for which pinched double roots exist.  Furthermore, we will observe that these singularities come in two varieties: either $G_\lambda$ is meromorphic in a neighborhood of the singularity or it fails to be analytic due to a branch point of the dispersion relation.

Recall (\ref{eq:Glambda}).  The first component, $G^{11}_\lambda$, satisfies the differential equation $(\mathcal{L}_u-\lambda)G^{11}_\lambda=\delta(y)$, for which it is readily computed that
\be G^{11}_\lambda(\xi-y)=\left\{ \begin{array}{c} \frac{1}{\nu_u^--\nu_u^+}e^{\nu_u^-(\lambda)(\xi-y)}\quad \text{for} \quad \xi>y \\
\frac{1}{\nu_u^--\nu_u^+}e^{\nu_u^+(\lambda)(\xi-y)}\quad \text{for} \quad \xi<y
\end{array}\right.. \label{eq:G11}\ee
In an analagous fashion,
\be 
G^{22}_\lambda(\xi-y)=\left\{ \begin{array}{c} \frac{1}{\nu_v^--\nu_v^+}e^{\nu_v^-(\lambda)(\xi-y)}\quad \text{for} \quad \xi>y \\
\frac{1}{\nu_v^--\nu_v^+}e^{\nu_v^+(\lambda)(\xi-y)}\quad \text{for} \quad \xi<y
\end{array}\right.. 
\label{eq:G22} 
\ee
It is easily verified that $G^{11}_\lambda$ has a singularity and loses analyticity at $\lambda=\alpha-\frac{s^2}{4d}$ while $G^{22}_\lambda$ does so at $\lambda=1-\frac{s^2}{4}$.  Selecting $s=s_{u,lin}=2\sqrt{d\alpha}$ places the singularity of $G^{11}_\lambda$ on the imaginary axis, while selecting $s=s_{v,lin}=2$ does the same for the $v$ component.  Note that the pinched double roots giving rise to the anomalous spreading speed have no effect on the analyticity of $G^{11}_\lambda$ or $G^{22}_\lambda$.  In other words, when $\beta=0$, the spreading speeds are just these linear spreading speeds and the pinched double root giving rise to the anomalous speed plays no role in the dynamics.

However, we are interested in the case of $\beta\neq 0$.  In this event, the non-diagonal element $G^{12}_\lambda$ is non zero and describes the linear behavior of the $u$ component due to coupling with the $v$ component.  We will see that the pinched double root $\nu_u^\pm=\nu_v^\mp$ may be relevant here.  To compute $G^{12}_\lambda$ we consider the linear equation
\[ (\mathcal{L}_u-\lambda)G^{12}_\lambda+\beta G^{22}_\lambda =0.\]
Skipping the specifics of the derivation, the general solution is computed on either half line $(\xi>y, \xi<y)$ to be
\be G^{12}_\lambda(\xi-y)=\left\{ \begin{array}{c} c_+(\lambda)e^{\nu_u^-(\lambda)(\xi-y)}-\frac{\beta}{\nu_u^--\nu_u^+}\frac{1}{d_u(\nu_v^-,\lambda)}e^{\nu_v^-(\lambda)(\xi-y)}\quad \text{for} \quad \xi>y \\ c_-(\lambda)e^{\nu_u^+(\lambda)(\xi-y)}-
\frac{\beta}{\nu_u^--\nu_u^+}\frac{1}{d_u(\nu_v^+,\lambda)}e^{\nu_v^+(\lambda)(\xi-y)}\quad \text{for} \quad \xi<y
\end{array}\right., \label{eq:g12} \ee
with the initial conditions enforced at $\xi=y$ so that $G^{12}_\lambda$ is continuously differentiable.  This prescribes the constants
\begin{eqnarray*}
c_+(\lambda)&=&\frac{\beta}{(\nu_u^--\nu_u^+)(\nu_v^--\nu_v^+)}\left(\frac{\nu_u^+-\nu_v^+}{d_u(\nu_v^+,\lambda)}-\frac{\nu_v^--\nu_v^+}{d_u(\nu_v^-,\lambda)}\right) \\
c_-(\lambda)&=& \frac{\beta}{(\nu_u^--\nu_u^+)(\nu_v^--\nu_v^+)}\left(\frac{\nu_v^+-\nu_u^-}{d_u(\nu_v^+,\lambda)}-\frac{\nu_v^--\nu_u^-}{d_u(\nu_v^-,\lambda)}\right).
\end{eqnarray*}

We make several immediate observations.  First, we see that $G^{12}_\lambda$ loses analyticity whenever any of the eigenvalues $\nu_{u,v}^\pm(\lambda)$ does.  To the right of the branch points $\lambda=\alpha-\frac{s^2}{4d}$ and $\lambda=1-\frac{s^2}{4}$ we then have that $G^{12}_\lambda$ is analytic with the possible exception of those $\lambda$ values for which $d_u(\nu_v^\pm,\lambda)=0$, or whenever a spatial eigenvalue for the $v$ subsystem, i.e. $\nu_v^\pm$, is equal to a spatial eigenvalue for the $u$ subsystem.  At any double root that does not satisfy the pinching condition, the pointwise Green's function has a removable singularity and therefore these points do not pose an obstacle to analytic continuation of $G_\lambda$.  On the other hand, pinched double roots where $\nu_u^\pm=\nu_v^\mp$ lead to poles of the pointwise Green's function.  The location of these poles can be found exactly as was done above, by computing those values of $\lambda$ for which the roots in (\ref{eq:nus}) form pinched double roots.  When $s=s_{anom}$ this pole lies at $\lambda=0$.  This establishes Theorem~\ref{thm:linearss}.

\begin{remark} 
Theorem~\ref{thm:linearss} demonstrates that linear spreading speeds are not continuous with respect to system parameters.  For parameters within the anomalous regime, the spreading speeds for $\beta>0$ are strictly larger than in the uncoupled case.  Note also that the particular value of $\beta$ is not important in so far as the linear spreading speed is concerned.  Based upon numerical simulations, we will make some conjectures as to the role of $\beta$ in section~\ref{sec:numerics}.
\end{remark}

\begin{remark}\label{rem:absolute}
Another perspective is obtained when considering the operator $\mathcal{L}$ posed on an exponentially weighted function space.  Namely, let $\eta>0$ and consider $L^2_\eta(\mathbb{R})=\left\{ \phi\in L^2(\mathbb{R}) \ | \ \phi(x)e^{\eta x}\in L^2(\mathbb{R})\right\}$.  Then $\mathcal{L}:H_\eta^2(\mathbb{R})\times H_\eta^2(\mathbb{R}) \to L_\eta^2(\mathbb{R})\times L_\eta^2(\mathbb{R})$  is isomorphic to $\mathcal{L}_\eta:H^2(\mathbb{R})\times H^2(\mathbb{R}) \to L^2(\mathbb{R})\times L^2(\mathbb{R})$, with
\[  \mathcal{L}_\eta=\left(\begin{array}{cc} d\partial^2_\xi+(s-2d\eta)\partial_\xi+(d\eta^2-s\eta+\alpha) & \beta \\
  0 & \partial^2_\xi+(s-2\eta)\partial_\xi+(\eta^2-s\eta+1) \end{array}\right). \]
The question then becomes whether for some fixed $s$, there exists a choice of $\eta>0$ so that the essential spectrum of $\mathcal{L}_\eta$ is pushed into the stable half plane.  This is related to the location of the absolute spectrum, see \cite{sandstede00}.  If the absolute spectrum is unstable, then no choice of exponential weight $\eta$ can stabilize the system.  In terms of the example considered here, the absolute spectrum consists of those values of $\lambda\in\mathbb{C}$ for which 
\[ \mathrm{Re} \  \nu_{u,v}^\pm(\lambda,s)=\mathrm{Re} \ \nu_{u,v}^\mp(\lambda,s).\]
Note that pinched double roots are elements of the absolute spectrum.  Consider parameter values in the anomalous regime.  Then, for any $s<s_{anom}$, there exists a pinched double root in the right half plane and no choice of $\eta>0$ will lead to decay in the weighted norm.  However, at $s=s_{anom}$ the pinched double root lies at $\lambda=0$ and all of the remaining absolute spectrum lies in the stable half plane.  Thus, for any $s\geq s_{anom}$, there exists an exponential weight so that the essential spectrum of $\mathcal{L}_\eta$ is pushed out of the open right half of the complex plane and pointwise decay is achieved in the weighted space.

\end{remark}

\section{Anomalous spreading speeds in the nonlinear system}~\label{sec:nonlinear}
We now turn our attention to spreading speeds in the nonlinear system.  For fixed values of $(d,\alpha)$, the linear spreading speed computed in section~\ref{sec:linear} gives a prediction for the spreading speed for the nonlinear system.  We will focus on parameter values for which the linear system predicts an anomalous spreading speed.   Numerical simulations, see section~\ref{sec:numerics}, suggest that this anomalous linear speed is only observed in the nonlinear system when $d<1$, see Figure~\ref{fig:parameterspace}.  In this section, we will investigate these two regimes and discuss possible mechanisms by which the nonlinear system adopts the linear speed in one parameter regime but not the other.

Anomalous linear spreading speeds arise due to poles of the pointwise Green's function induced by pinched double roots wherein $\nu_u^\pm=\nu_v^\mp$.  We will prove that double roots with $\nu_u^-=\nu_v^+$ are {\em irrelevant} in that the spreading speed observed in the nonlinear problem is slower than the anomalous linear speed induced by the pole.  On the other hand, we will examine the {\em relevant} case where $\nu_u^+=\nu_v^-$ and argue that these poles lead to faster nonlinear spreading through a the adoption of weaker decay rates by the $u$ component.  

We remark that in both the relevant and irrelevant case we are dealing with pulled fronts wherein the the front propagation is being driven by the instability ahead of the front interface.  For pulled fronts, one can glean a great deal of insight into the nonlinear dynamics from the linearization about the unstable state and we will make repeated use of properties of the linearized system in this section.

\paragraph{The scalar Fisher-KPP equation}

We begin with some basic facts concerning the nonlinear, scalar Fisher-KPP equation,
\be u_t=du_{xx}+\alpha(u-u^2).\label{eq:FKPPscalar} \ee
We emphasize the role of nonlinear decay rates.  Nonlinear traveling fronts, $U_{KPP}(x-st)$ can be constructed for all positive wavespeeds as heteroclinic trajectories in the phase space of the traveling wave ordinary differential equation, $dU''+sU'+\alpha (U-U^2)=0$.  The spatial decay rate of this front is the asymptotic rate of decay of the heteroclinic orbit to the origin.  This depends on the eigenvalues of the origin.  We see a connection with the linear theory of section~\ref{sec:linear}, since these eigenvalues are the roots $\nu_u^\pm(0,s)$ from (\ref{eq:nus}).  For $s<2\sqrt{d\alpha}$, these eigenvalues are complex conjugate and oscillatory decay is observed for $U_{KPP}(x-st)$.  For $s>2\sqrt{d\alpha}$, there are two possible decay rates corresponding to the two branches $\nu_u^\pm$.  It is known that for the Fisher-KPP equation, the nonlinear traveling front always approaches the origin along the eigenspace corresponding to the weaker of the two decay rates.  For $s=2\sqrt{d\alpha}$ the origin is a degenerate node and the nonlinear front has algebraic decay.

According to the marginal stability criterion, see \cite{dee83}, the front selected from compactly supported initial data is the unique front for which perturbations of the front profile neither grow nor decay exponentially when viewed pointwise in a frame moving with the speed of the front.  Denote by $\mathcal{L}_{KPP}$ the linearization of the Fisher-KPP equation about the nonlinear traveling front $U_{KPP}(x-st)$.  That is,
\[ \mathcal{L}_{KPP}= d \partial_\xi^2 +s\partial_\xi +\alpha (1-2U_{KPP}(\xi)).\]
The spectrum of this operator in $L^2(\mathbb{R})$ is unstable.  The front is said to be marginally stable if there exists an exponentially weight $W(\xi)=1+e^{\eta \xi}$ with $\eta>0$ for which the spectrum of the linearized operator posed on $L^2_W(\mathbb{R})=\{\phi\in L^2(\mathbb{R})| W(\xi)\phi(\xi)\in L^2(\mathbb{R})\}$ can be pushed to the imaginary axis, but no $\eta$ exists for which the spectrum can be pushed completely into the stable half plane.  

Given a Fisher-KPP front propagating with speed $s$, there are two obstacles to marginal stability: the absolute spectrum and resonance poles or embedded eigenvalues.  Recall from \cite{sandstede00} (see also Remark~\ref{rem:absolute}) that the location of the absolute spectrum demarcates how far the essential spectrum of the operator $\mathcal{L}_{KPP}$ can be pushed with the aid of exponential weights.  The following is known
\begin{itemize}
\item For $s>2\sqrt{d\alpha}$, the nonlinear traveling fronts are spectrally stable for certain choices of exponential weights, see \cite{sattinger}.  The derivative of the front, $\partial_\xi U_{KPP}(\xi)$, is not a zero eigenvalue since it has weak spatial decay and does not lie in the weighted space $L^2_W$.  These fronts are inaccessible from compactly supported initial data, although initial data with the same weak exponential decay will lead to traveling fronts with this speed.  
\item For $s=2\sqrt{d\alpha}$, the nonlinear traveling front is marginally stable, see \cite{kirchgassner92}.  This front is selected by compactly supported initial data and is a pulled front since the nonlinearly selected speed is equal to the linear spreading speed.
\item For $s<2\sqrt{d\alpha}$, the fronts are unstable for any exponential weight due to the instability of the absolute spectrum.
\end{itemize}
There is one more point to keep in mind.  Consider positive initial data for (\ref{eq:FKPPscalar}) consisting of a compactly supported perturbation of a Heaviside step function.  As $t$ tends to infinity, the solution of this initial value problem converges in $L^\infty(\mathbb{R})$ to the marginally stable Fisher-KPP traveling front\footnote{with a logarithmic correction to the wavespeed, see \cite{bramson83}}.  However, we emphasize that for any fixed value of $t>0$, $u(t,x)$ converges to zero as $x\to\infty$ at a rate faster than any exponential.  This will be important in what follows.

\begin{figure}[ht]
\centering
   \includegraphics[width=0.4\textwidth]{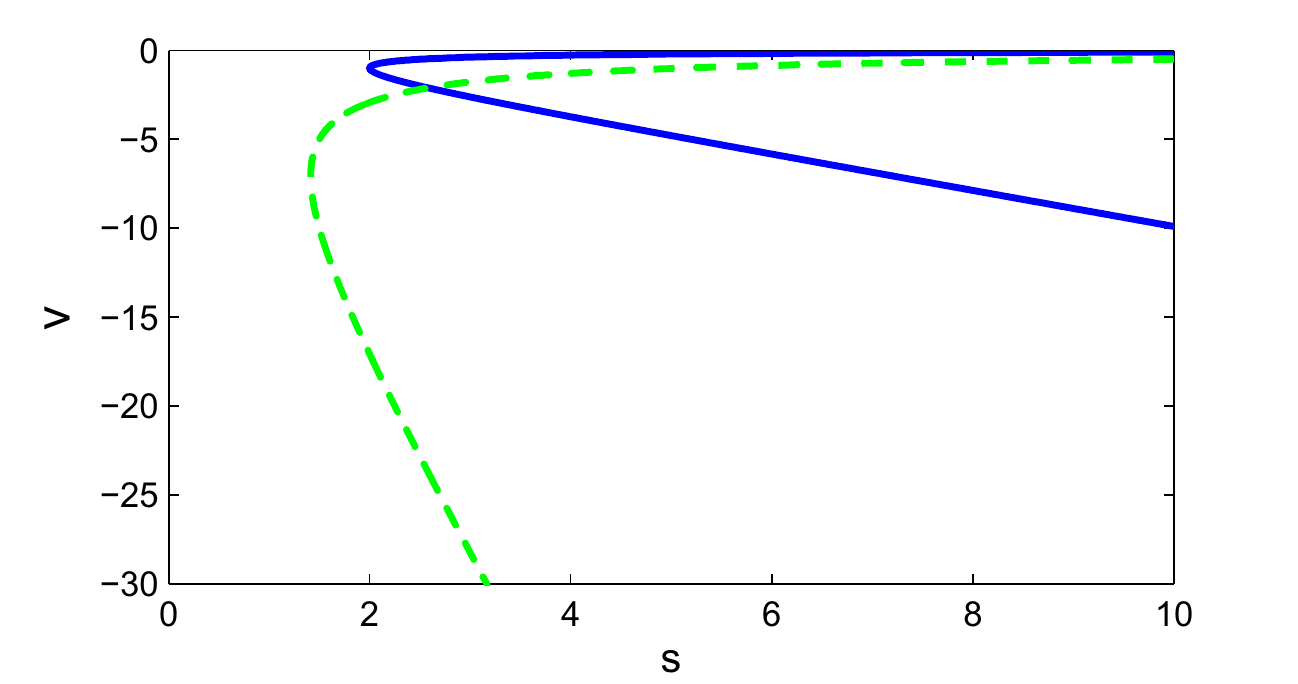}
   \includegraphics[width=0.4\textwidth]{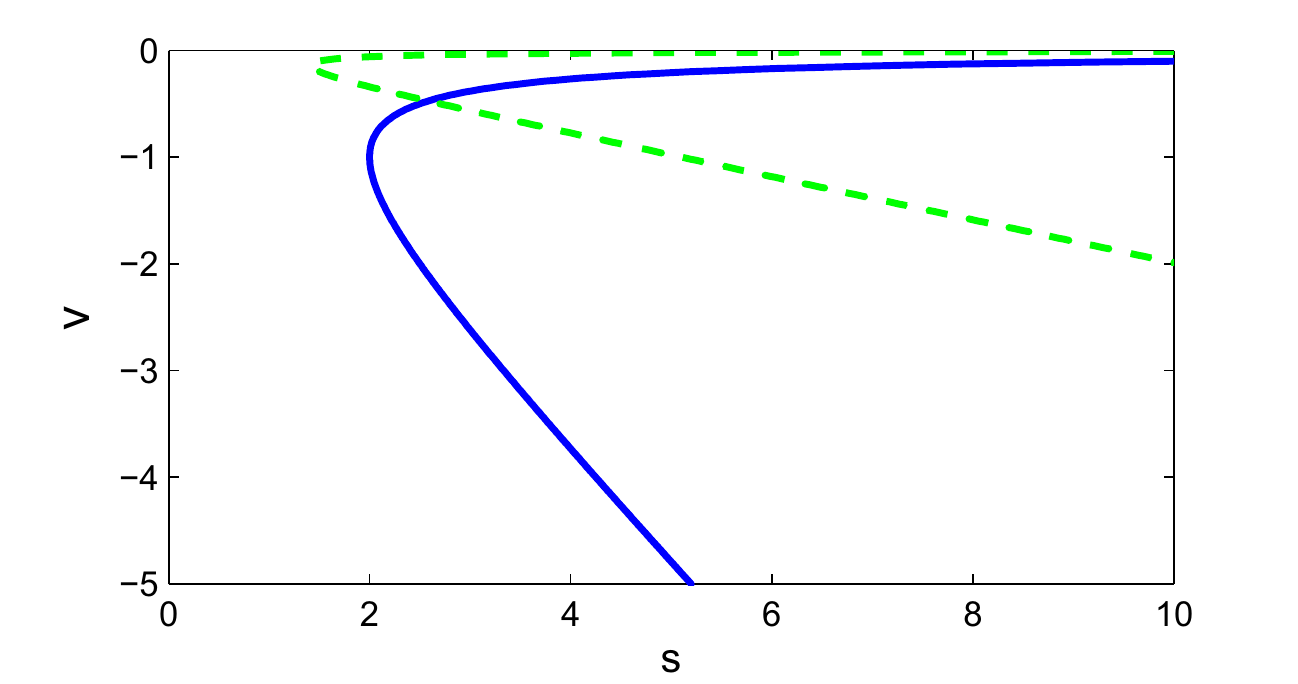}  
\caption{Plots of the group velocity $s_g$ as a function of the spatial decay rate $\nu$.  The solid line is this curve for the $v$ component.  Note the branch point at the linear spreading speed of $2$ with selected spatial decay rate of $-1$.  The dashed line is the decay rate -- group velocity plot for two different values of $(\lambda,\nu)$.  On the left is the curve for $d=0.1$ and $\alpha=5$.  Note the pinched double root that occurs involving $\nu_u^+$ and $\nu_v^-$.  This pinched double root leads to faster observed speeds of propagation in both the linear and nonlinear systems.  On the right is the curve for $d=5$ and $\alpha=0.1$.  Here, the system exhibits anomalous spreading for the linear system but not for the nonlinear system.}
\label{fig:maintool}
\end{figure}

\paragraph{Group velocities and nonlinear decay rates}
We will now discuss group velocities for the linearized system.  We will emphasize their relationship with nonlinear decay rates and the linear spreading speed.  Roots of the dispersion relation with $\lambda=0$, i.e. $\nu_{u,v}^\pm(0,s)$, will reappear here.  We begin the discussion by considering the linearized equation for the $u$ component in isolation,

\[ u_t=du_{xx}+\alpha u.\]
As we did when deriving the dispersion relation above, take the ansatz $u(t,x)=e^{\lambda t}e^{\nu x}$ and substitute this into the linear equation.  Assuming that $\nu\in\mathbb{R}$ and $\nu<0$, then we find that $\lambda=d\nu^2+\alpha$.  From this the group velocity,
\be s_{g}=-\frac{\lambda(\nu)}{\nu}=-d\nu-\frac{\alpha}{\nu},\label{eq:groupvelocity}\ee
gives the speed with which this exponential propagates in the linear equation.   This relationship can be visualized in $(\nu,s)$ space, see Figure~\ref{fig:maintool}.  For each value of $s>2\sqrt{d\alpha}$ there are two possible decay rates that propagate with speed $s$ in the linear system.  These values of $\nu$ are denoted $\nu_u^\pm$ and called spatial eigenvalues.  These eigenvalues are exactly the eigenvalues of the fixed point at the origin for the traveling wave equation,
\[ du_{\xi\xi}+su_\xi+\alpha u-\alpha u^2=0.\]
In this way, for every wavespeed $s>2\sqrt{d\alpha}$, the group velocity equation relates this speed to the two possible decay rates of nonlinear traveling fronts.  The upper branch corresponds to traveling front profiles with weak decay.  As was mentioned previously, these fronts are stable in an exponentially weighted space, but are not accessible from compactly supported initial data.  On the other hand, if the initial data is not compactly supported but instead has an exponential decay rate belonging to this weak branch then the initial data will evolve into a traveling front moving with the speed prescribed by the group velocity calculation above.  


\paragraph{Spreading in the coupled system}
The same general picture holds for the $v$ component.  The group velocities for both equations in isolation can be plotted together, see Figure~\ref{fig:maintool}.  Anomalous linear speeds exist when the upper branch of one of these curves intersects the lower branch of the other.  Numerical simulations suggest that nonlinear anomalous speeds are only observed when the double root is relevant (i.e. involves $\nu_u^-$ and $\nu_v^+$).  We now come to discuss mechanisms that lead to these anomalous speeds and argue as to why they persist to the nonlinear regime in some cases but not in others.  This will draw on the relation between pinched double roots, group velocities and nonlinear decay rates.


\begin{itemize}
\item {\bf Relevant double roots -- adoption of weak decay by the $u$ component} -- 

Consider $d$ and $\alpha$ in the anomalous linear regime for which the associated pinched double root involves $\nu_u^-$ and $\nu_v^+$.  These parameters lie in the lobe labeled $A$ in Figure~\ref{fig:parameterspace} with $d<1$.  Recall that nonlinear traveling fronts exist that travel with speeds greater than the linear speed $2\sqrt{d\alpha}$, but these fronts are not accessible from compactly supported initial data.  

We hypothesize the following mechanism that leads to anomalous spreading in the nonlinear system.  The $v$ component evolves from its initial condition and converges towards a traveling front solution.  This solution is coupled to the $u$ component through an inhomogeneous term.  The effect of this term is to establish decay rates in the $u$ component that are similar to those in the $v$ component.

If these decay rates are sufficiently weak, then the previously inaccessible weak fronts with speed $s>2\sqrt{d\alpha}$ are able to be observed in the $u$ component despite compactly supported initial data.  This leads to faster observed speeds of propagation.  Recall that the solution $v(t,x)$ is converging to the Fisher-KPP front with speed two and exponential decay $\xi e^\xi$.  However,  ahead of the front interface the solution $v(t,x)$ converges to zero faster than $e^{-x}$ for any fixed time.  We emphasize that the mechanism leading to faster propagation speeds is more subtle than simply the adoption of the weak decay rate associated to the Fisher-KPP front $V_{KPP}(x-2t)$. Faster speeds of propagation are observed only if there exists a double root at $\nu_u^+=\nu_v^-$ and the observed spreading speed is then the linear spreading speed associated to this double root.  In general, this speed is slower than the spreading speed associated to weak fronts with decay $e^{-x}$.  This is due to the rather complicated route that compactly supported initial data take in the convergence to their asymptotic form of a traveling front solution. 

Rigorous validation that this mechanism leads to anomalous nonlinear speeds of propagation remains an open problem.  One might be tempted to adopt a dynamical systems approach and seek traveling front solutions.  The dynamics of the $v$ component evolve towards a traveling front solution and it might therefore seem natural to study invasion speeds in the inhomogeneous scalar equation,
\be u_t=du_{xx}+\alpha(u-u^2)+\beta V_{KPP}(x-2t)(1-u).\label{eq:usingtheKPP} \ee
As we alluded to above, taking this approach will over-estimate the actual speed of propagation.  The effect of the inhomogeneity in (\ref{eq:usingtheKPP}) is to impart the decay rate of the KPP front onto the $u$ component.  The selected speed of propagation for (\ref{eq:usingtheKPP}) can then be determined by the group velocity calculation in (\ref{eq:groupvelocity}).  This speed will be larger than the speed selected by compactly supported initial data.  

Since (\ref{eq:main}) obeys a maximum principle, one might hope to construct compactly supported sub-solutions that would validate the existence of the nonlinear anomalous spreading speed.
Constructing such a sub-solution would require some detailed knowledge concerning the route that the $v$ component takes during its convergence to the traveling front $V_{KPP}(\xi)$.  This does not appear to be completely understood and is beyond the scope of this paper.




\item {\bf Irrelevant double roots -- weak growth and fast diffusion in the $u$ equation} -- 

When $d>1$, anomalous linear speeds arise due to pinched double roots wherein $\nu_v^-=\nu_u^+$.   Since it is $\nu_v^-$ that is involved, we do not expect the mechanism described when $d<1$ to lead to faster speeds in the nonlinear regime.   As was noted above, $v(t,x)$ is converging to a traveling front with speed two and asymptotic spatial decay rate $-1>\nu_v^-$. Recall also that the root $\nu_v^-$ corresponds to weak spatial decay of the $v$ component.  Faster rates of decay are observed ahead of the front interface, but weaker decay rates are not.  In this way, the pinched double root leading to anomalous linear speeds does not lead to anomalous speeds in the nonlinear regime.  For the particular case studied here of (\ref{eq:main}), this will be made rigorous in Theorem~\ref{thm:nofastspeed}.

Before proceeding to the theorem, we pause to explain the physical mechanism that leads to anomalous linear speeds and argue as to why this mechanism can not survive in the nonlinear regime.  That anomalous spreading speeds are observed in the linearized equation is apparent from the pointwise Green's function.  The physical mechanism leading to this phenomena in the linear regime can be best conceptualized in the limit of large diffusion in the $u$ component.  Here, the $v$ component grows pointwise at an exponential rate.  The $v$ solution enters into the $u$ equation as an inhomogeneous source term.  The $u$ dynamics adopt the pointwise exponential growth of the $v$ component, but with now a much larger constant of diffusion.  This leads to faster linear spreading speeds.  Note that this was the same mechanism discovered in \cite{holzerLV} to describe anomalous linear spreading in the Lotka-Volterra competition model, (\ref{eq:lotka}).  Observe that this mechanism depends critically on the exponential in time growth of the $v$ component.  For the nonlinear system, such growth occurs only transiently and therefore anomalous linear spreading does not persist for the nonlinear system in the irrelevant case.  
\end{itemize}

We will prove that irrelevant double roots do not induce anomalous nonlinear spreading for (\ref{eq:main}).  We consider initial data $0\leq u(0,x)\leq 1$, $0\leq v(0,x)\leq 1$ both compactly supported perturbations of the Heaviside step function $H(-x)$.  Define the invasion point 
\[\delta(t)=\max_{x\in\mathbb{R}} \left\{ u(t,x)=\frac{1}{2}\right\}.\]
The spreading speed is defined as,
\[ s_{sel}=\limsup_{t\to\infty} \frac{\delta(t)}{t}.\] 
We will construct explicit super-solutions, see for example \cite{fife77}, that will establish the following result.
\begin{thm}\label{thm:nofastspeed} Consider the initial value problem (\ref{eq:main}), with initial conditions, $0\leq u(0,x)\leq 1$, $0\leq v(0,x)\leq 1$, both compactly supported perturbations of Heaviside step functions $H(-x)$.  Suppose that $d>1$ and $\alpha<\frac{d}{2d-1}$ and let $s_{anom}$ be given as in Theorem~\ref{thm:linearss}.  Then 
\[ s_{sel}\leq \max\{2,2\sqrt{d\alpha}\}.\]
\end{thm}
\begin{pf}
Consider $\max\{2,2\sqrt{d\alpha}\}<s<s_{anom}$.  We will construct a front-like super-solution that propagates to the right with speed $s$.  To begin, it is known that for any $s>2$ and $C_v>0$,  
\[\bar{v}(t,x)=\min\{1,C_ve^{\nu_v^-(0,s)(x-st)}\}\]
is a super-solution for the $v$ dynamics.  Moreover, given $v(0,x)$ as above there exists a $C_v>0$ so that $\bar{v}(0,x)>v(0,x)$ and therefore the spreading speed of the $v$ component is bounded above by $s$.  The super-solution $\bar{v}(t,x)$ places an upper bound on the spatial decay rate at any fixed time $t>0$.
This is a refinement of the statement above that the solution of the Fisher-KPP initial value problem with compactly supported initial data converges to zero faster than any exponential, see \cite{ebert00}.  

We will now turn our attention to the $u$ component and construct a similar super-solution for any $\max\{2,2\sqrt{d\alpha}\}<s<s_{anom}$.  Consider
\[ N(u)=u_t-du_{xx}-\alpha u -\beta v (1-u) +\alpha u^2.\]
We seek $\bar{u}(t,x)$ so that $N(\bar{u})\geq 0$ for all $t>0$ and $x\in\mathbb{R}$.  Let
\be\bar{u}(t,x)=\left\{ \begin{array}{ccc} 1 & \text{for} & (x-st)\leq \tau \\
 C_ue^{\nu_u^-(x-st)}+C_v\kappa e^{\nu_v^-(x-st)} & \text{for} & (x-st)>\tau \end{array} \right. .\label{eq:bu} \ee
For any $C_v>0$, we will find $C_u^*(C_v)>0$ and $\tau(C_u,C_v)$ so that $\bar{u}(t,x)$ is a super-solution for any $C_u>C_u^*$.  One easily verifies that $u=1$ is a super solution.  If $\tau$ is taken to be larger than $\tau_{min}:=(\nu_v^-)^{-1}\log C_v$, then for $x-st>\tau$ we have that $N(\bar{u})$ is equivalent to
\[ N(\bar{u})=\bu_t-d\bu_{xx}-\alpha\bu-\beta C_ve^{-\nu_v^-(x-st)} +\beta (\bar{v}-v)+\beta v\bu+\alpha \bu^2.\]
Consider
\[ \bu_t=d\bu_{xx}+\alpha \bu +\beta C_ve^{-\nu_v^-(x-st)}.\]
This equation has the solution,
\[ \tilde{u}(t,x)=C_ue^{\nu_u^-(x-st)}+C_v\kappa e^{\nu_v^-(x-st)},\]
with
\[ \kappa=\frac{-\beta}{d(\nu_v^-)^2+s\nu_v^-+\alpha}<0.\]
For $(d,\alpha)$ under consideration, it is also the case that $\nu_v^-<\nu_u^-$.  This implies that for any $C_u>0$,  $\tilde{u}(t,x)>0$ for $x$ sufficiently large.  Also $\tilde{u}(t,x)$ has a unique maximum at  
\[ \xi_{max}=-\frac{\log(\frac{-C_u\nu_u^-}{C_v\kappa \nu_v^-})}{\nu_u^--\nu_v^-},\]
for $\xi=x-st$.  Fix $C_u^*>0$ sufficiently large so that both $\xi_{max}>\tau_{min}$ and the value of $\bu(t,x)$ at this maximum exceeds one.  Then for any $C_u>C_u^*$, let $\tau(C_u,C_v)$ be such that $\tilde{u}(t,\tau+st)=1$.  Then $\bu(t,x)$ in (\ref{eq:bu}) is continuous.  It remains to verify that $N(\bu(t,x))>0$ for $x-st>\tau$.  On this semi-infinite interval we have,
\[ N(\bu(t,x))=N(\tilde{u}(t,x))=\beta (\bar{v}-v)+\beta \tilde{u} v +\alpha \tilde{u}^2>0,\]
establishing $\bu(t,x)$ as a super-solution. 

Since $0\leq u(0,x)\leq 1$ is assumed to be a compactly supported perturbation of the Heaviside step function $H(-x)$, we can find $C_u(u(x,0))>C_u^*$ so that $\bu(0,x)\geq u(0,x)$.  Then $\bu(t,x)\geq u(t,x)$ and we find that $s_{sel}\leq s$.  This construction holds for any $\max\{2,2\sqrt{d\alpha}\}<s<s_{anom}$.  This establishes the theorem.

\end{pf}

\section{A stable system inducing anomalous spreading}~\label{sec:stablebutfast}
Viewing the evolution of (\ref{eq:mainlinearized}) in a frame of reference traveling at the anomalous spreading speed, we observe that the $v$ component is convectively stable.  That is, for any fixed $\xi=x-s_{anom}t$ the $v$ component converges pointwise exponentially fast in $t$ to the zero state.  Taking this point of view, we expect that anomalous spreading will be observed when the zero state is pointwise stable for the $v$ dynamics and not just convectively stable.  

\begin{figure}[ht]
\centering
   \includegraphics[width=0.4\textwidth]{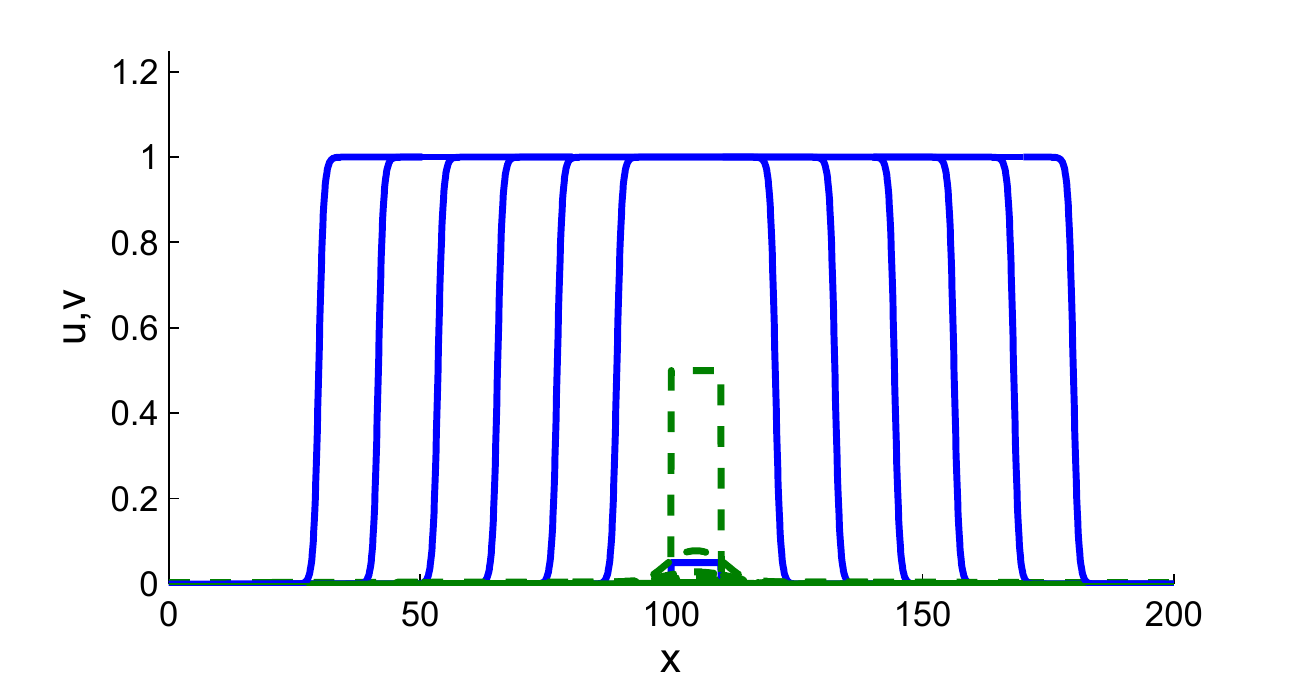} 
   \includegraphics[width=0.4\textwidth]{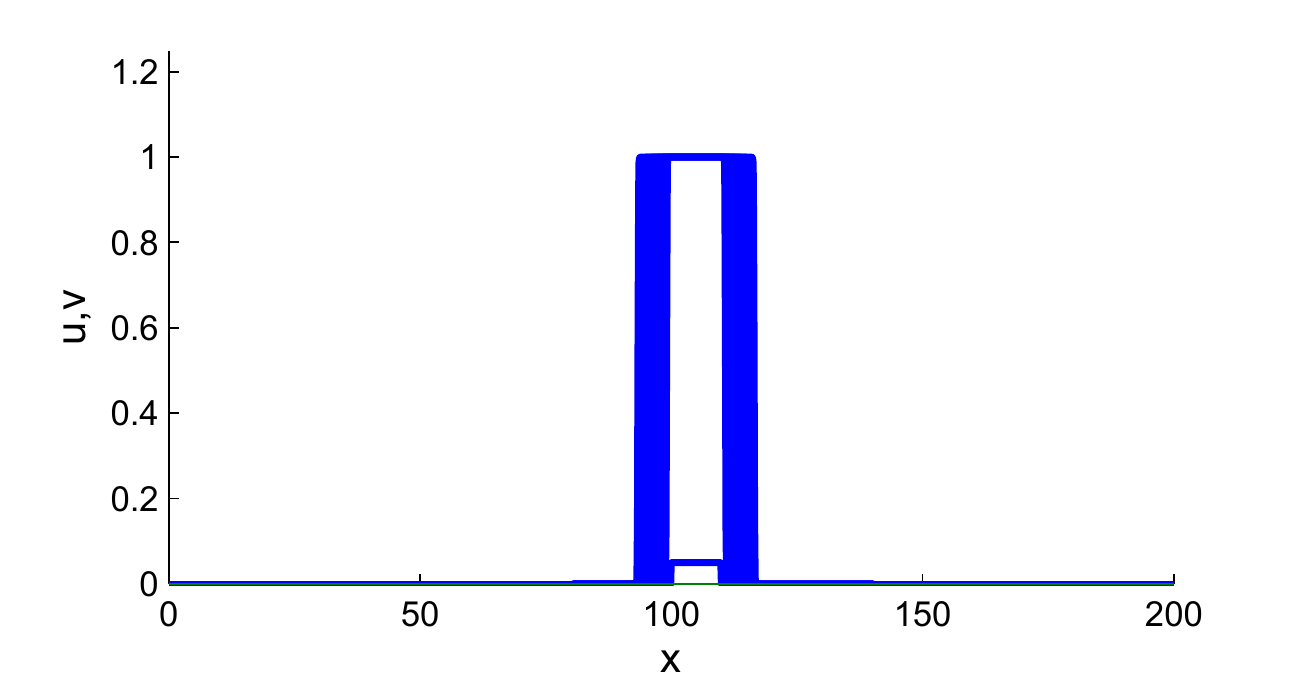} 
\caption{Anomalous spreading for (\ref{eq:stablebutfast}) with $d=0.001$, $\alpha=4$, $\beta=2$ and $\gamma=0.1$.  The solid line is the $u$ component while the dashed line is the $v$ component.  Note that the $v$ component relaxes to zero while the $u$ component spreads.  The evolution is shown at times $t=0,6,12,18,24,30,36$.  On the left the evolution is shown for the full system.  On the right, the same simulation is run but with identically zero initial conditions for the $v$ component.  The spreading speed in the absence of coupling is $2\sqrt{0.004}\approx .1265$ whereas the anomalous speed is approximately $1.9765$.}
\label{fig:stablebutfast}
\end{figure}

We provide the following example,
\begin{eqnarray}
u_t &=& du_{xx}+\alpha\left(u-u^2\right)+\beta v (1-u) \nonumber \\
v_t &=& v_{xx}- \gamma v \label{eq:stablebutfast}.
\end{eqnarray}
We again consider compactly supported, positive perturbations of the homogeneous zero state.  With this initial data, $v(t,x)$ will decay to zero pointwise exponentially fast for all $x$ as $t\to\infty$.  With $v(t,x)$ converging to zero, one might expect that the $u$ component will then spread with the speed $2\sqrt{d\alpha}$.  However, this is not the case as one can observe in Figure~\ref{fig:stablebutfast}.  The decay rates of $v(t,x)$ as $x\to\pm\infty$ play an important role in the dynamics despite the fact that the solution is very small and converging to zero.  This should be contrasted with the dynamics where $v$ is set identically equal to zero.  Here the $u$ component converges, as expected, to the Fisher-KPP front traveling with speed $2\sqrt{d\alpha}$.  

This is completely analogous to the unstable case described above.
As before, the linear spreading speed can be determined from the dispersion relation.  The spatial eigenvalues for the $v$ component are now
\[ \nu_v^\pm=\frac{-s}{2}\pm\frac{1}{2}\sqrt{s^2+4\gamma+4\lambda}.\]
The evolution of the linearized dynamics are again prescribed by a pointwise Green's function and computing pinched double roots leads to a precise characterization of the linear spreading speed.  In particular, we find that poles of the pointwise Green's function arise whenever $\nu_v^-=\nu_u^+$ and for the wavespeed
\[ s=\sqrt{\frac{\alpha+\gamma}{1-d}}-\gamma\sqrt{\frac{1-d}{\alpha+\gamma}}.\]
This anomalous speed exists for the parameter values
\[ d<\frac{1}{2}\ \ \text{and} \ \ \alpha>\frac{d\gamma}{1-2d}.\]

We have the same dichotomy here between relevant and irrelevant double roots.  When the double root involved is $\nu_v^-=\nu_u^+$, then the anomalous linear spreading speed is observed in the nonlinear regime.  In example (\ref{eq:stablebutfast}), this is the only possible type of double root.  Irrelevant double roots occur when coupling is done in the opposite direction.  That is, if the identical equations were coupled in the opposite direction then this pinched root would once again exist and give rise to an anomalous spreading speed in the linear equation, but would not lead to faster speeds of propagation in the nonlinear system.  This is what occurs in (\ref{eq:lotka}) and explains why the fast linear spreading that was observed in that model does not persist to the nonlinear regime, see \cite{holzerLV}.

\section{Numerical simulations}~\label{sec:numerics}
\begin{figure}[ht]
\centering
   \includegraphics[width=0.37\textwidth]{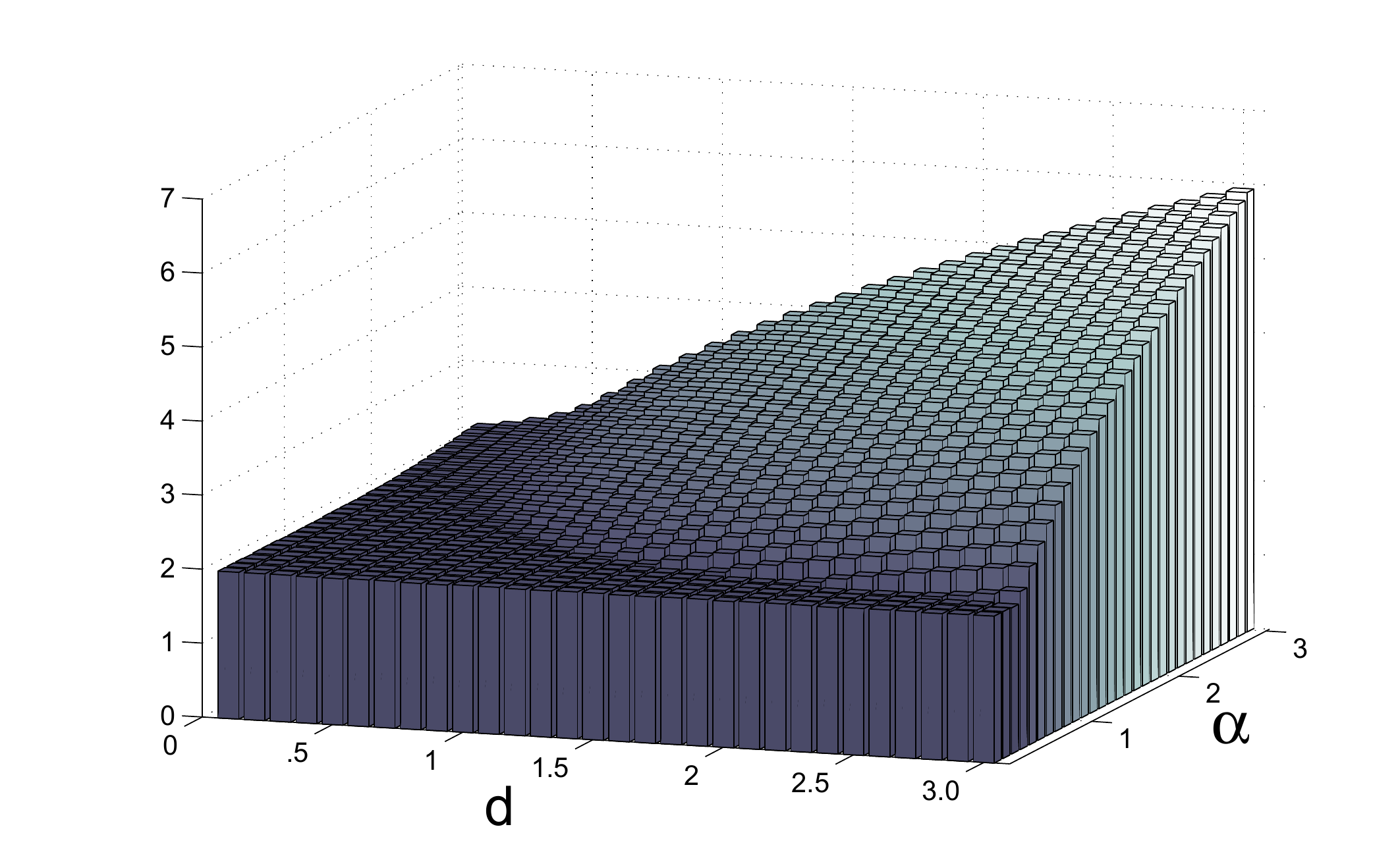}
   \includegraphics[width=0.4\textwidth]{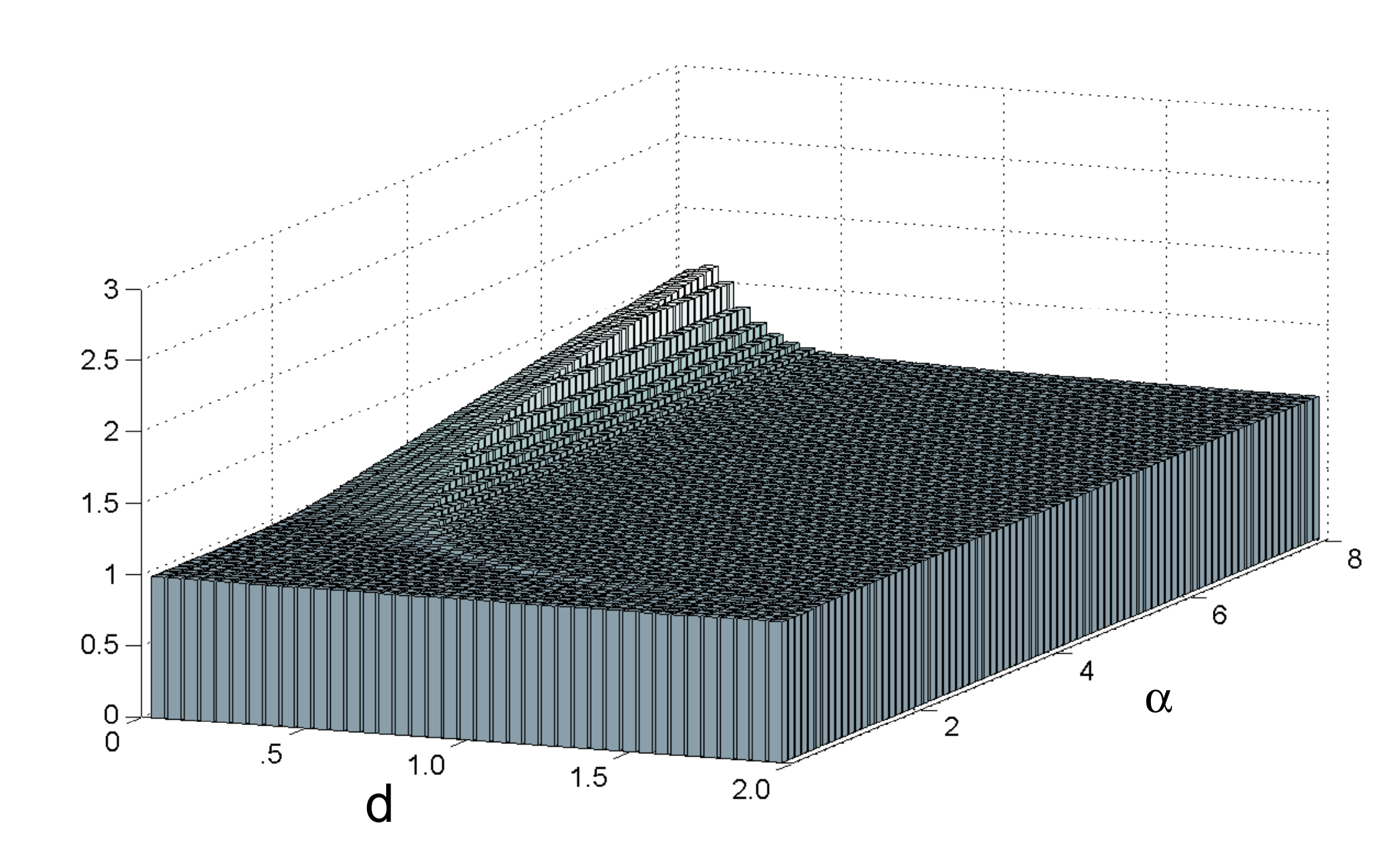}  
\caption{On the left, we plot numerically observed spreading speeds for the $u$ component for $d\in [0,3]$ and $\alpha\in [0,3]$.  Note that the spreading speed remains two in the linearly anomalous regime for $d>1$.  On the right, we plot the relative error $s_{observed}/\max\{2,2\sqrt{d\alpha}\}$.  }
\label{fig:33square}
\end{figure}

\begin{figure}[ht]
\centering
   \includegraphics[width=0.4\textwidth]{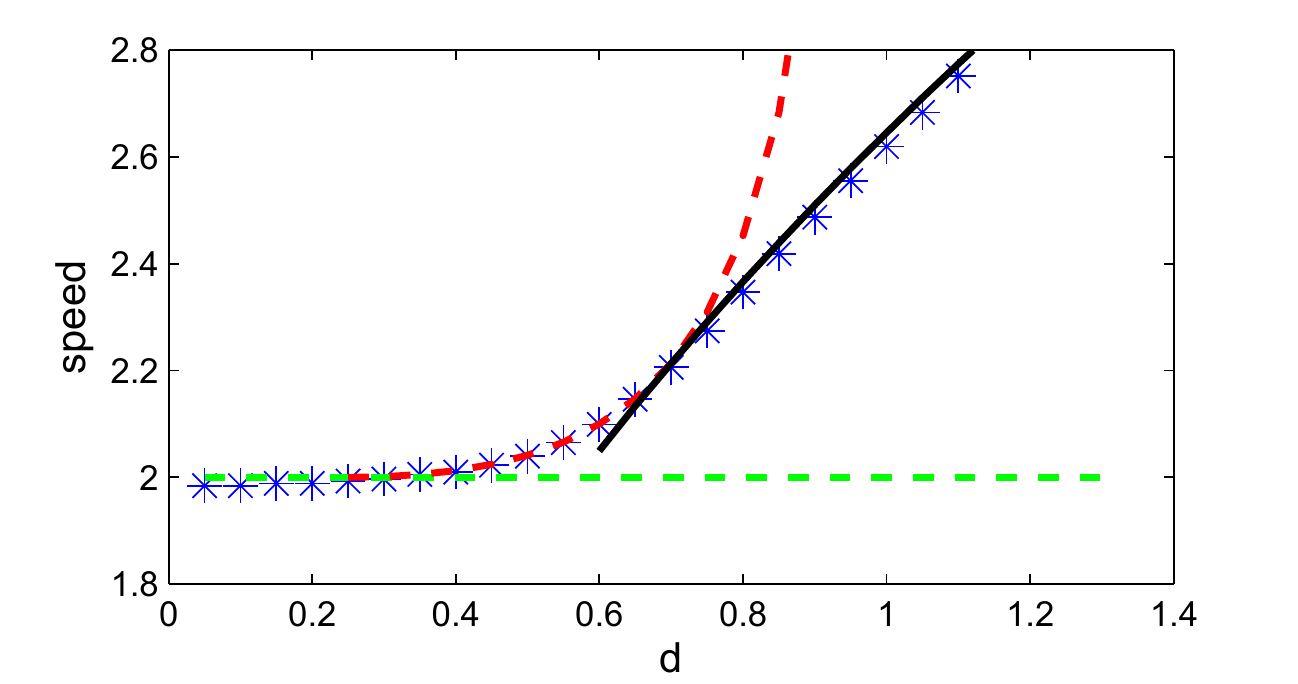}
   \includegraphics[width=0.4\textwidth]{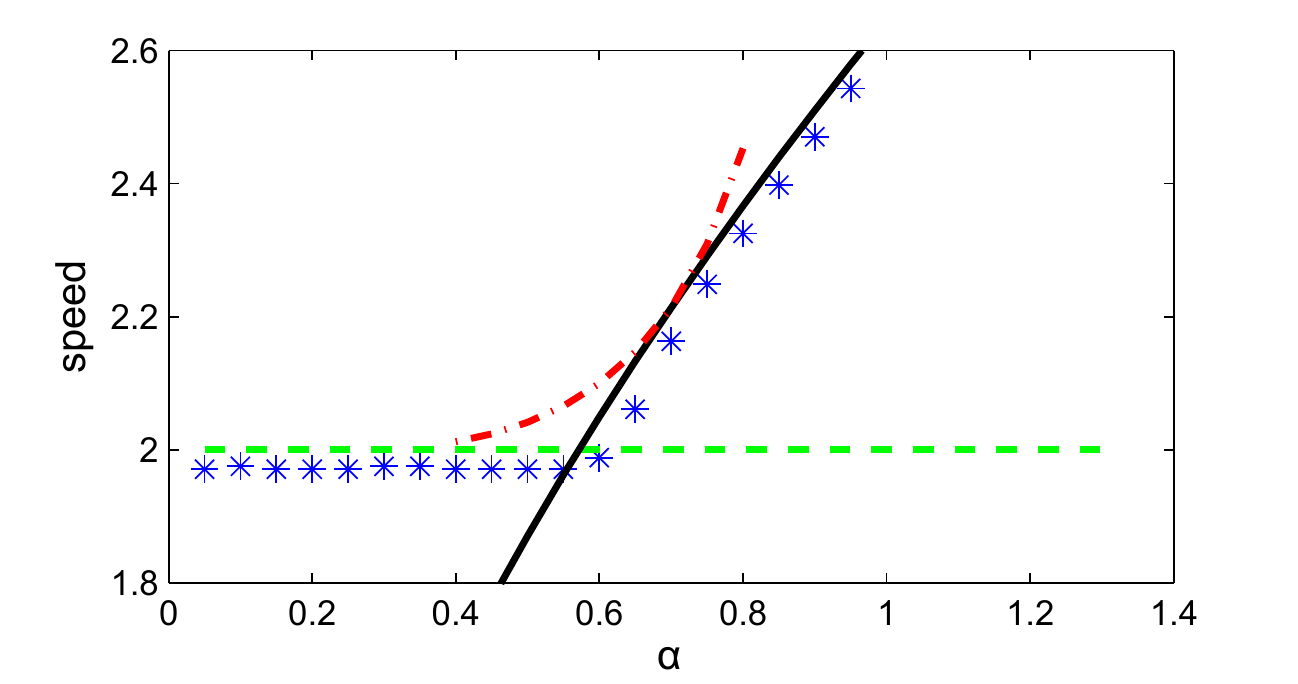}  
\caption{Here we compare numerically derived spreading speeds within two separate sections of the $(d,\alpha)$ parameter plane.  On the left, we consider $\alpha=1.75$ and vary $d$ between $0$ and $1.4$.  The $*$ are numerically derived spreading speeds while the curved dashed line is the anomalous speed, the solid line is the speed $s_{u,lin}$ and the straight dashed line is $s_{v,lin}$.  On the right, we do the same comparison, but now with $d=1.75$ and $\alpha$ varied.  As predicted, these parameters lie in the region where there is no anomalous spreading and slower invasion speeds are observed.    }
\label{fig:sectionfig}
\end{figure}

In this final section, we compare numerically derived spreading speeds for the full nonlinear system to the linear spreading speeds predicted in Theorem~\ref{thm:linearss}.  Recalling our discussion in section~\ref{sec:nonlinear}, we expect to observe anomalous nonlinear spreading only when $d<1$ and within the parameter regime detailed in Theorem~\ref{thm:linearss}.  We compute spreading speeds using a Crank-Nicolson type finite difference method.  Spatial domains of varying sizes were considered with Neumann boundary conditions imposed.  Step function initial data in both the $u$ component and the $v$ component was selected. Spreading speeds were repeatedly computed  over short time scales.  Following an initial transient, the computed spreading speed relaxed to a nearly constant value.  These numerically observed wavespeeds for the $u$ component are presented in Figure~\ref{fig:33square} and Figure~\ref{fig:sectionfig}.

\begin{figure}[ht]
\centering
   \includegraphics[width=0.31\textwidth]{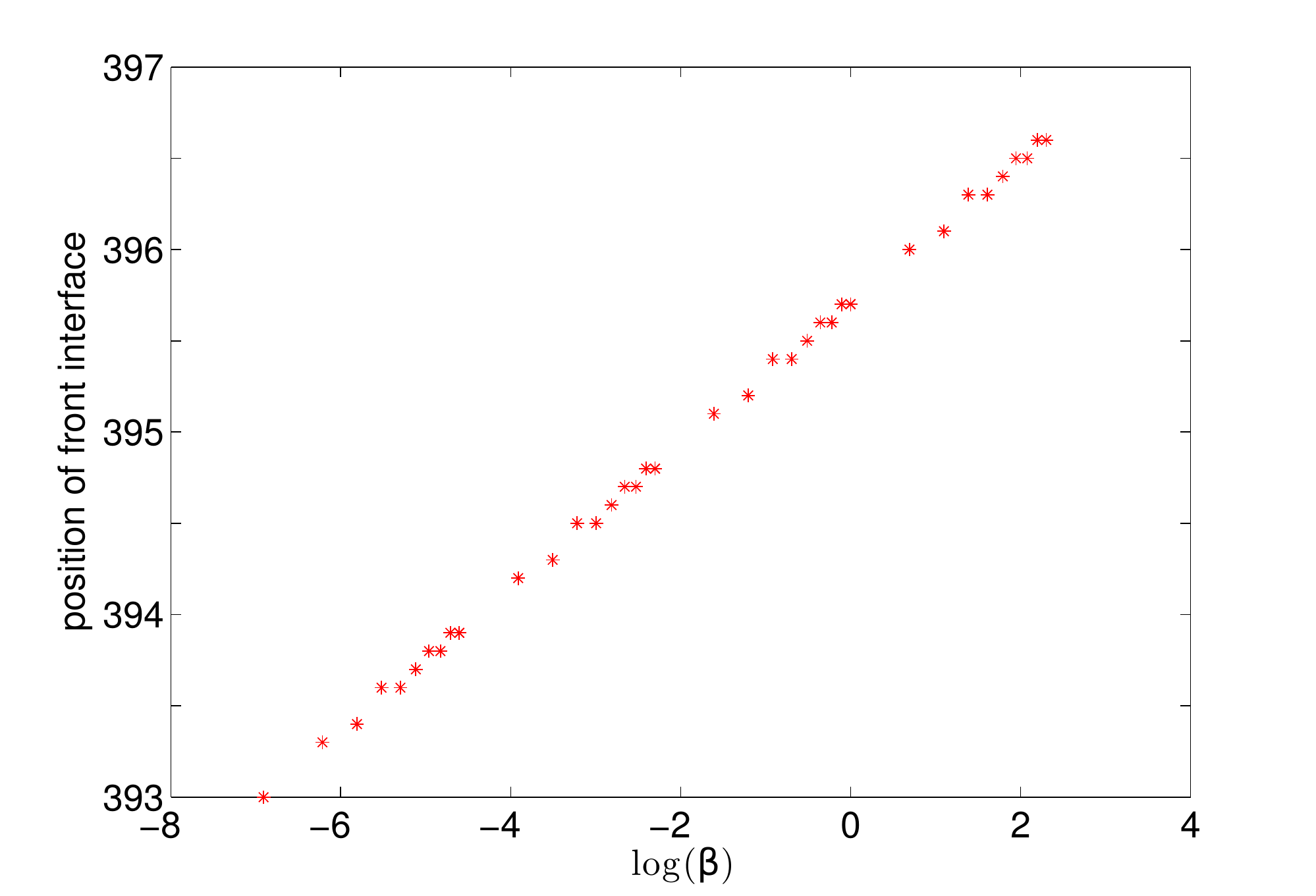}
   \includegraphics[width=0.33\textwidth]{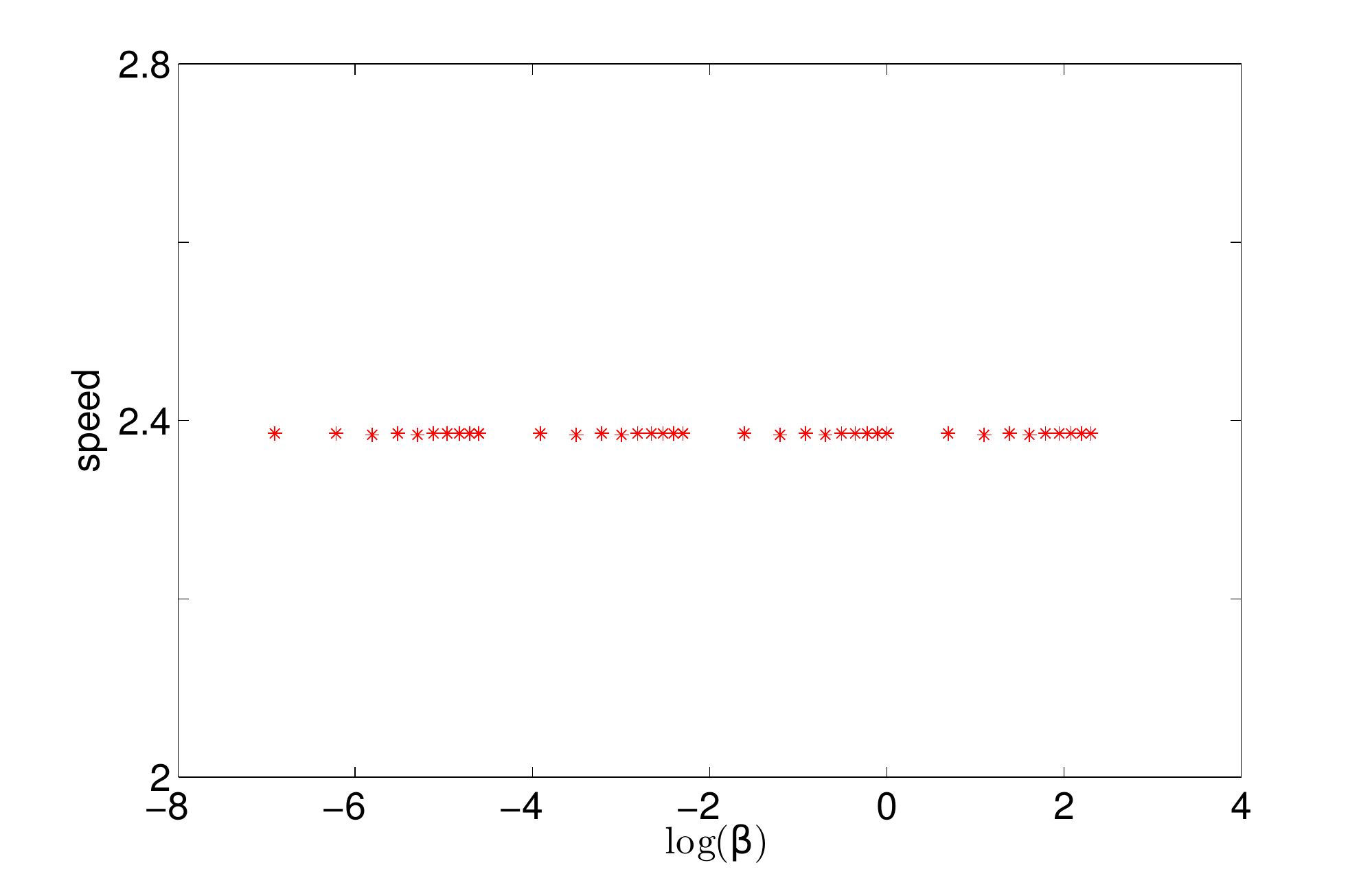} 
   \includegraphics[width=0.31\textwidth]{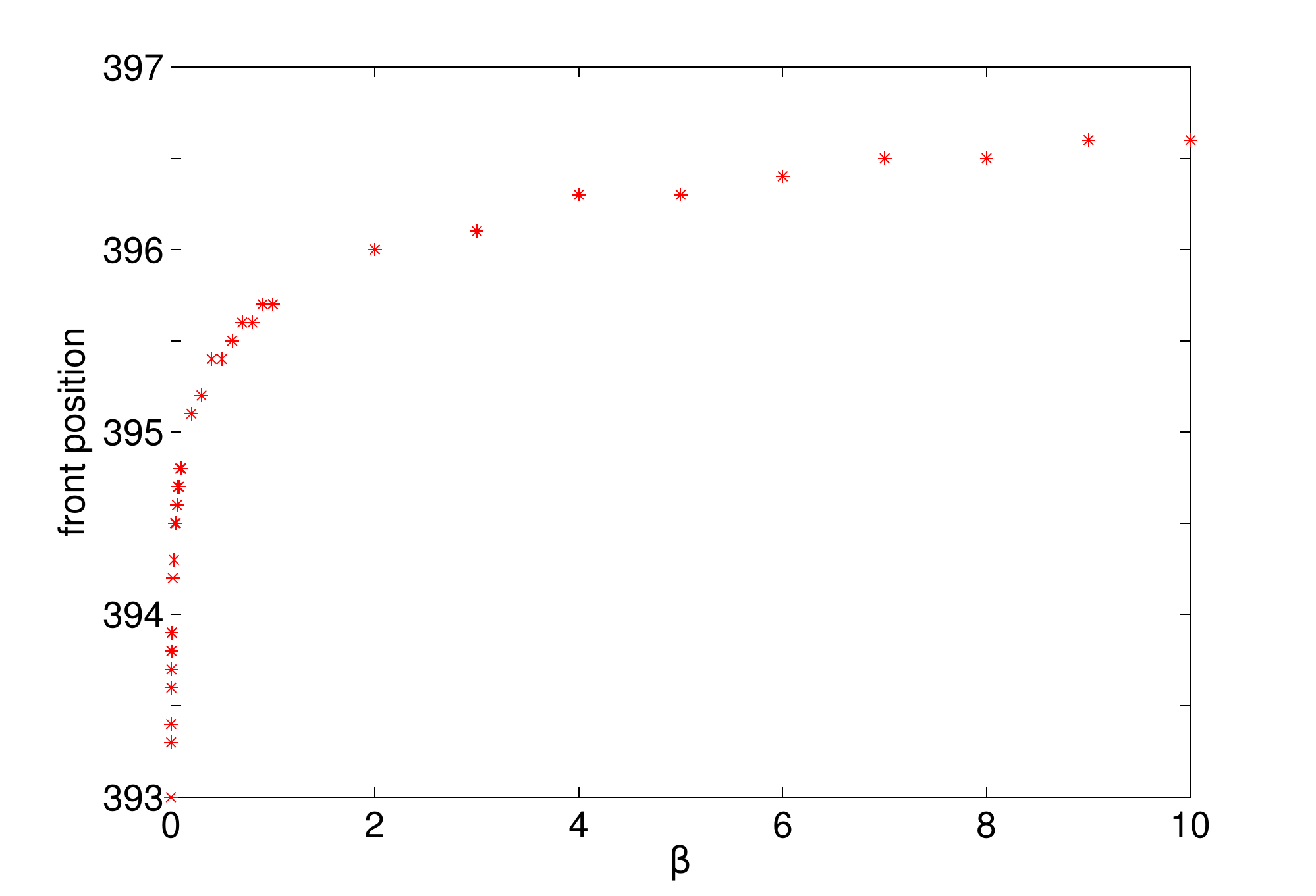} 
\caption{The results of numerical simulations of (\ref{eq:main}) for various values of $\beta>0$.  Here we have fixed the parameters $d=0.1$ and $\alpha=7$.  In the left two panels, we start from identical step function initial data and simulate the system on the domain $[0,1000]$ for $t\in[0,100]$.  We compare the speed and location of the front interface as a function of the logarithm of $\beta$.  We observe that, as expected, the speed does not depend on the particular value of $\beta$ under consideration.  However, the left panel suggests a linear relationship between the position of the front interface and the logarithm of $\beta$.     }
\label{fig:betarole}
\end{figure}

We have thus far largely ignored the role of the parameter $\beta$ in the dynamics of system (\ref{eq:main}).  We will make some qualitative observations based upon numerical simulations. Theorem~\ref{thm:linearss} shows that the value of $\beta>0$ does not play a role in the determination of the linear spreading speed.  The same appears to be the case in the nonlinear system.  Here, we conjecture that $\beta$ does not influence the speed selection but instead plays a role in determining which particular translate of the traveling front that is selected.  Numerical simulations corroborate this view, see Figure~\ref{fig:betarole}.   Note that a logarithmic relationship between the position of the traveling front and the value of $\beta>0$.  This is reminiscent of the scalar Fisher-KPP equation with weakly decaying initial data.  There, if initial data with weak decay $e^{-\gamma x}$ converges to a particular translate of a traveling front, then initial data with decay $\beta e^{-\gamma x}$ will converge to the same traveling front but shifted in space by $\gamma^{-1}\log\beta$.

\begin{remark}The mechanism hypothesized in section~\ref{sec:nonlinear} as an explanation for anomalous spreading depends upon the decay rates of the $v$ component.  As such, numerical roundoff error will eventually lead to slower observed spreading speeds in the nonlinear system than the anomalous ones.  This is more pronounced when the invasion occurs into a non-zero unstable state.  However, even when it is the zero state that is being invaded far ahead of the $v$ front interface underflow eventually occurs and the $u$ component is no longer able to resolve the decay rates of the $v$ solution.  This leads to eventual slowing of the $u$ component in numerical simulations.
\end{remark}

\section{Discussion}~\label{sec:discussion}
In this article, we have investigated anomalous spreading speeds in a system of coupled Fisher-KPP equations.  We conclude with a brief commentary on more general aspects of this phenomena as well as a discussion of unresolved issues.

The fact that the $v$ component completely decouples in the nonlinear system (\ref{eq:main}) is not a prerequisite for anomalous spreading.  For example, the nonlinear system
\begin{eqnarray*}
u_t &=& du_{xx}+\alpha\left(u-u^2\right)+\beta v (1-u) \nonumber \\
v_t &=& v_{xx}+\epsilon uv(1-u)+\left(v-v^2\right),
\end{eqnarray*}
exhibits nonlinear anomalous spreading where the $v$ component travels at speed $2$ and the $u$ component travels at the speed selected in the partially decoupled ($\e=0$) case.  We note that qualitatively similar invasion phenomena has been observed in \cite{bell09} in the context of an ecological model of the interaction of healthy and diseased red and grey squirrels.  This model is much more complicated than (\ref{eq:main}), but numerical simulations reveal behavior reminiscent of anomalous spreading.

One might also be interested in the behavior of the system under some small perturbation that destroys the skew-product nature of the linearization.  For example, consider the system
\begin{eqnarray}
u_t &=& du_{xx}+\alpha\left(u-u^2\right)+\beta v (1-u) \nonumber \\
v_t &=& v_{xx}+\epsilon u(1-u)+\left(v-v^2\right). \label{eq:epsperturb}
\end{eqnarray}
When $\e=0$, we recover (\ref{eq:main}).  However, for $\e\neq 0$, the linearization about the unstable state no longer has the requisite skew-product structure.  In this case, the pinched double root of the dispersion relation that led to a pole of the pointwise Green's function perturbs to a branch point wherein $G_\lambda$ loses analyticity.  These branch points are then dynamically relevant irregardless of whether they corresponded to nonlinearly relevant anomalous spreading speeds in the $\e=0$ case.  That is, for $\e$ small and positive the observed spreading speeds for (\ref{eq:epsperturb}) are small perturbations of the linear spreading speeds given in Theorem~\ref{thm:linearss} and plotted in the left panel of Figure~\ref{fig:parameterspace}.  If $\e<0$, then positivity is not preserved for (\ref{eq:epsperturb}) and solutions diverge.  However, if the quadratic nonlinear terms are replaced with cubic ones then for $\e<0$ the linear spreading speed corresponds to values of $\lambda^*$ for which $\mathrm{Im}(\lambda^*)\neq 0$.  In this regime, the invasion front forms patterns in its wake with long wavelengths.  See \cite{goh11} for a precise example of this phenomena in the context of a phase-field model.  In this way, system (\ref{eq:main}) lies on the boundary between systems with invasion fronts that leave patterns in their wake and those that leave homogeneous states.

To conclude, we remark that a rigorous mathematical description of the anomalous invasion process remains an open question.  As we pointed out above, the use of a traveling front approach is complicated by several factors.  Most glaringly, is that the dynamics of the full system evolve on two separate time scales and therefore no choice of the traveling wave coordinate reduces the problem to an ordinary differential equation.  More subtle, is the fact that the observed spreading speed depends on the ``transient'' evolution of compactly supported initial data towards a traveling front profile.

\section*{Acknowledgments}  The author thanks Jonathan Sherratt for useful conversations and for pointing the author towards his work in \cite{bell09}.  In addition, the author acknowledges Arnd Scheel for several insightful discussions.
This research was supported by the NSF (DMS-1004517).

\bibliographystyle{abbrv}
\bibliography{MHmaster}

\end{document}